\documentclass[journal,comsoc]{IEEEtran}

\usepackage{authblk}
\usepackage{algorithmic}
\usepackage{graphicx} 
\usepackage{xcolor}
\usepackage{textcomp}
\usepackage{multirow}
\usepackage{graphicx}
\usepackage{mathtools}  
\usepackage{amsmath}
\usepackage{amssymb}
\usepackage{tabulary}
\usepackage{booktabs}
\usepackage[ruled,linesnumbered]{algorithm2e}
\usepackage[utf8]{inputenc}
\usepackage{comment}
\usepackage{subfig} 
\usepackage{float}
\usepackage{adjustbox}
\emergencystretch=\maxdimen
\usepackage[cmintegrals]{newtxmath}
\usepackage{enumitem}
\usepackage{bm}
\usepackage{cite}
\usepackage{rotating}

\begin{document}
\title{Multi-Objective Offloading Optimization in MEC and Vehicular-Fog Systems: A Distributed-TD3 Approach}

\author{Frezer~Guteta~Wakgra, 
        Binayak~Kar, 
        Seifu~Birhanu~Tadele,
        Shan-Hsiang~Shen,
        and~Asif~Uddin~Khan

\thanks{F.G. Wakgra, B. Kar, S.B. Tadele, and S.-H. Shen are with the Department of Computer Science and Information Engineering, National Taiwan University of Science and Technology, Taipei 106, Taiwan (e-mail: haftiyam@gmail.com, bkar@mail.ntust.edu.tw, seife.brhan@gmail.com, and sshen@csie.ntust.edu.tw).}
\thanks{A.U. Khan is with the School of Computer Engineering, KIIT Deemed to be University, Bhubaneswar, Odisha, India. (E-mail: asif.khanfcs@kiit.ac.in)}
}

\maketitle

\begin{abstract}

The emergence of 5G networks has enabled the deployment of a two-tier edge and vehicular-fog network. It comprises Multi-access Edge Computing (MEC) and Vehicular-Fogs (VFs), strategically positioned closer to Internet of Things (IoT) devices, reducing propagation latency compared to cloud-based solutions and ensuring satisfactory quality of service (QoS). However, during high-traffic events like concerts or athletic contests, MEC sites may face congestion and become overloaded.
Utilizing offloading techniques, we can transfer computationally intensive tasks from resource-constrained devices to those with sufficient capacity, for accelerating tasks and extending device battery life. In this research, we consider offloading within a two-tier MEC and VF architecture, involving offloading from MEC to MEC and from MEC to VF.
The primary objective is to minimize the average system cost, considering both latency and energy consumption. To achieve this goal, we formulate a multi-objective optimization problem aimed at minimizing latency and energy while considering given resource constraints. To facilitate decision-making for nearly optimal computational offloading, we design an equivalent reinforcement learning environment that accurately represents the network architecture and the formulated problem.
To accomplish this, we propose a Distributed-TD3 (DTD3) approach, which builds on the TD3 algorithm. Extensive simulations, demonstrate that our strategy achieves faster convergence and higher efficiency compared to other benchmark solutions.

 
\end{abstract}


\section{Introduction}
\IEEEPARstart{T}{he} main concept of the Internet of Things (IoT) is that various devices comprising different technologies, such as industrial actuators, wearable devices, autonomous vehicles, smart sensors, etc., will be connected and communicate with each other without human intervention through the Internet \cite{intro_01}. It is a key paradigm in creating a smart service and making an informed decision for monitoring, control, and management purposes \cite{intro_02}. IoT devices generate a massive amount of data which requires a huge amount of computational power and storage capacity. However, its limited computational capacity and battery power pose a big challenge in meeting the quality of service (QoS) \cite{intor_03}. To address these issues, an offloading concept can be used \cite{kar2023offloading}. Offloading involves migrating resource-intensive computations from a resource-constrained device to a device with adequate resources to speed up computations and save battery life. IoT devices can offload their tasks to cloud computing, which offers abundant computing and storage capacity. However, cloud computing has its own drawbacks, such as a single point of failure, lack of location awareness, reachability, and latencies \cite{intro_05}.

Multi-access Edge Computing (MEC) has been introduced by the European Telecommunications Standards Institute (ETSI) as a supplement to cloud computing and mobile edge computing by deploying edge servers at base stations \cite{kar2023cost}. It serves as an effective method to solve the limited resources problems in IoT devices \cite{intro_06}. In MEC environments, servers with less processing capabilities than cloud-based servers are positioned closer to IoT devices on the edge of the cellular network, enabling them to provide computing services for IoT devices. IoT devices can optimize their functionality by transferring computational tasks to Mobile Edge Computing (MEC) sites \cite{wu2014}. This process involves leveraging device-to-device (D2D) communication through wireless networks to improve their performance \cite{zhao2017}. 
Once the tasks have been completed, the IoT devices will receive the final results back. In this way, idle network resources can be utilized efficiently, and QoS can be enhanced by reducing the energy consumption of IoT devices and the latency in carrying out computational operations \cite{intro_07}. 

In the context of MEC, high-traffic scenarios, such as those occurring during sporting events or music concerts, can potentially overload an MEC site \cite{intro_11}. To address this, horizontal and vertical federations can be established. Horizontal federation facilitates the offloading of a portion of traffic to neighboring MEC sites \cite{lin2023cost}, while vertical federation involves offloading traffic into the connected Vehicle Fog (VF) \cite{lin2019cost}. In an Internet of Vehicles (IoV) ecosystem, vehicles equipped with sensors, communication devices, and smart technologies communicate with each other, infrastructure, and central control systems \cite{agbaje2022survey}. Vehicles with limited computational capacity can collaborate to serve as both a communication and computation platform. Additionally, a connected device handling numerous tasks could be parking lot vehicular fog or traffic intersection vehicular fog, contributes to further enhancing overall performance \cite{intro_13}.

Over the past decade, researchers have demonstrated significant interest in computational offloading. 
In an MEC system, computational offloading is classified into issues concerning the control plane and the management plane. The control plane swiftly addresses incoming traffic, while the management plane predicts future traffic or task arrival rates based on historical data \cite{kar2023offloading}. 
However, existing research on this federated architecture has primarily concentrated on the management plane. 
The offloading problem is typically regarded as a component of the management plane, involving upward vertical offloading from a UE or vehicles to an MEC or cloud server and horizontal offloading between vehicles, MECs, or fogs, with the UE or vehicles making those decisions \cite{dai2020,mukherjee2019}. In this study, we aim to bridge this gap by delving into the control plane within the same architecture. 
The offloading problem is considered a component of the MEC control plane, involving horizontal offloading to the neighboring MEC site or downward vertical offloading to the VF, where the offloading decisions are made by the MEC. 
It's worth noting that only a limited number of studies have explored this particular area \cite{intro_19}, to the best of our current knowledge.
\color{black}

In this study, we focus on investigating horizontal offloading (from MEC to MEC) and vertical offloading (from MEC to VFs) in a two-tier MEC and VF architecture. When IoT devices generate a substantial volume of traffic, it's categorized as hotspot traffic and routed to the MEC site to harness its enhanced computation capabilities. However, the MEC site faces the challenge of potential overload due to its limited computational capacity, leading to significant computational delays that violate QoS requirements. To address this issue, incoming traffic can either be processed locally or offloaded to a nearby horizontally federated MEC site or a vertically federated VF to enhance response times. However, once traffic is offloaded to VFs, it undergoes rapid processing with reduced latency due to the extensive distributed computational capacity of VFs, albeit at the expense of increased energy consumption. Achieving a balance between traffic processing delay and energy consumption is imperative, with the primary objective being the simultaneous minimization of both average system latency and energy consumption.

Offloading decisions in this study are made using a deep reinforcement learning (DRL) agent, the distributed twin-delayed deep deterministic policy gradient (DTD3) algorithm, deployed at the control plane. The algorithm's main objective is to make an optimal offloading decision that simultaneously minimizes latency and maximizes battery life for the VFs while satisfying QoS requirements. Once the traffic is offloaded to the preferred site and computations are completed, the results are returned to the MEC site. In short, the main contributions of this work are summarized as follows:

\begin{enumerate}
\item We formulate an optimization problem of computational offloading for MEC and VF architecture using queueing theory, with the objective of minimizing the average weighted sum cost of the entire system in terms of latency and energy consumption.
\item We develop an equivalent reinforcement learning (RL) environment for the MEC and VF network and transform the above problem into it.
\item We propose an efficient DRL-based Distributed-TD3 algorithm to optimize offloading decisions for solving the problem.
\item We conduct extensive simulations to assess the efficiency of the RL environment and evaluate the performance of the proposed approach.
\end{enumerate}

The rest of the study is organized as follows. Section II describes the related work. Section III discusses the system modeling and problem formulation. Section IV introduces the proposed solution. Section V presents the parameter setting and result analysis. Finally, Section VI provides the conclusion of this work.

\section{Related Works}

Most previous research \cite{rw_01, rw_02, rw_03, rw_04, rw_05, rw_06, rw_08, rw_09, rw_10, zhou2022, budhiraja2022} primarily focused on designating the destination for offloaded tasks rather than quantifying the volume of traffic to be offloaded. Moreover, existing studies mainly explored the option for UE or vehicles to perform upward vertical offloading \cite{rw_01, rw_02, rw_03, rw_04, rw_05, rw_06, rw_08, rw_09, rw_10, zhou2022, budhiraja2022} to MEC or the cloud or horizontal offloading \cite{rw_01, rw_08, rw_10, zhou2022, budhiraja2022} to nearby vehicles or MECs. In contrast, this study addresses the challenge of relieving an overloaded MEC site, which can horizontally offload specific incoming tasks to neighboring MEC sites or vertically downward to a VF. The Distributed-TD3 algorithm is utilized to determine the optimal offloading decision, specifying both the location and quantity of tasks to be offloaded by an MEC site. 
This section provides a summary of literature studies that have explored different target networks for computational offloading, aiming to achieve diverse objectives using various RL techniques, as illustrated in Table \ref{tab:rw}.


\begin{table*}[htbp!]
  \centering
  \caption{Survey on offloading in different target networks using RL }
  \label{tab:rw}%
  \renewcommand{\arraystretch}{1.5}
  \resizebox{\textwidth}{!}{ %
    \begin{tabular}{|p{1cm}|p{1cm}|p{2cm}|p{1cm}|p{1cm}|p{1cm}|p{1cm}|p{1cm}|p{1cm}|p{1cm}|p{3cm}|p{1.5cm}|p{2cm}|}
        \hline
        \multirow{2}[4]{*}{\textbf{Papers}} & \multicolumn{1}{c|}{\multirow{2}[4]{*}{\textbf{Target Network}}} & \multicolumn{2}{p{10em}|}{\textbf{Target Plane}} & \multicolumn{2}{p{10em}|}{\textbf{Offloading Direction}} & \multicolumn{2}{p{10em}|}{\textbf{Offloading Type}} & \multicolumn{2}{p{10em}|}{\textbf{Fog Type}} & \multirow{2}[4]{*}{\textbf{Objective}} & \multirow{2}[4]{*}{\textbf{Constraints}} & \multirow{2}[4]{*}{\textbf{Approaches}} \\
        \cline{3-10}    \multicolumn{1}{|c|}{} &       & \textbf{Management} & \textbf{Control} & \textbf{Vertical} & \textbf{Horizontal} & \textbf{1 to 1} & \textbf{1 to n} & \textbf{Static } & \textbf{Dynamic } & \multicolumn{1}{c|}{} & \multicolumn{1}{c|}{} & \multicolumn{1}{c|}{} \\
         \hline
            \cite{rw_01}  & \multicolumn{1}{c|}{V $\leftrightarrow$ V, V $\uparrow$ RSU}  & $\checkmark$    & $\times$     & $\checkmark$ \space    & $\checkmark$      & $\checkmark$     & $\times$     & $\checkmark$     & $\times$    & Maximize QoE & Caching and Resource & DDPG \\
         \hline
            \cite{rw_02}  & \multicolumn{1}{c|}{V $\uparrow$ VEC}     & $\checkmark$     & $\times$     & $\checkmark$ $\space$      & $\times$     & $\checkmark$     & $\times$     & $\times$     & $\checkmark$     & Minimize Latency and Energy & Resource & PPO \\
         \hline
            \cite{rw_03}  & \multicolumn{1}{c|}{V $\uparrow$ E, V $\uparrow$ C}     & $\checkmark$     & $\times$     & $\checkmark$ \space     & $\times$     & $\checkmark$     & $\times$     & $\checkmark$     & $\times$     & Minimize Latency and Reliability & Latency and Reliability & RL \\
         \hline
            \cite{rw_04}  &  \multicolumn{1}{c|}{UE $\uparrow$ MEC}     & $\checkmark$     & $\times$     & $\checkmark$ \space      & $\times$     & $\checkmark$     & $\times$     & $\checkmark$     & $\times$     & Minimize Energy & Latency and Reliability & Q-learning Double-DQL\\
         \hline
            \cite{rw_05}  & \multicolumn{1}{c|}{V $\uparrow$ MEC}     & $\checkmark$     & $\times$     & $\checkmark$  \space    & $\times$     & $\checkmark$     & $\times$     & $\times$     & $\checkmark$     & Minimize Latency and Energy & Latency and Resource & Q-learning and multi-agent\\
         \hline
            \cite{rw_06}  & \multicolumn{1}{c|}{UE $\uparrow$ VEC}      & $\checkmark$     & $\times$     & $\checkmark$ \space     & $\times$     & $\checkmark$     & $\times$     & $\checkmark$     & $\times$     & Maximize the long-term utility of the network & Latency and Resource & Q-learning and DRL \\
         \hline
            \cite{rw_08}  & \multicolumn{1}{c|}{V $\leftrightarrow$ V, V $\uparrow$ RSU}     & $\checkmark$     & $\times$     & $\checkmark$ \space      & $\checkmark$     & $\checkmark$     & $\times$     & $\times$     & $\checkmark$     & Minimize Energy & Latency & Edmonds-Karp and DRL \\
         \hline
            \cite{rw_09}  & \multicolumn{1}{c|}{V $\uparrow$ MEC, V $\uparrow$ C}     & $\checkmark$     & $\times$     & $\checkmark$ \space      & $\times$     & $\checkmark$     & $\times$     & $\times$     & $\checkmark$     & Minimize Latency and Energy & Resource & DQN-based JCOTM \\
         \hline
            \cite{rw_10}  & \multicolumn{1}{c|}{V $\leftrightarrow$ V, V $\uparrow$ RSU}    & $\checkmark$   & $\times$     & $\checkmark$ \space      & $\checkmark$     & $\checkmark$     & $\times$     & $\times$     & $\checkmark$     & Maximize Capacity & Resource & Q-Leering based CCSRL\\
         \hline
         \cite{zhou2022}  & \multicolumn{1}{c|}{UE $\uparrow$ E, E $\leftrightarrow$ E, E $\uparrow$ C}    & $\checkmark$   & $\times$     & $\checkmark$ \space      & $\checkmark$     & $\checkmark$     & $\times$     & $\times$     & $\checkmark$     & Minimize Energy & Caching, Latency and Resource & DDPG\\
         \hline
         \cite{budhiraja2022}  & \multicolumn{1}{c|}{UE $\leftrightarrow$ VEC, UE $\uparrow$ MEC}    & $\checkmark$   & $\times$     & $\checkmark$ \space      & $\checkmark$     & $\checkmark$     & $\times$     & $\times$     & $\checkmark$     & Minimize Latency and Energy & Latency & SAC\\
         \hline
            Ours  & \multicolumn{1}{c|}{MEC $\leftrightarrow$ MEC, MEC $\downarrow$ VF}    & $\times$     & $\checkmark$       & $\checkmark$ \space      & $\checkmark$     & $\times$     & $\checkmark$     & $\times$     & $\checkmark$     & Minimize Latency and Energy & Resource & Distributed-TD3 \\
        \hline
    \end{tabular}  %
 }
\end{table*}%

He et al. \cite{rw_01} proposed an edge-enabled IoV designed to offload task-data chunks to maximize the quality of experience for vehicles. They introduce a deep deterministic policy gradient (DDPG)-based DRL algorithm to achieve this goal. The study considers vehicle caching, computation, and the edge server's channel as constraints. Zhan et al. \cite{rw_02} modeled the offloading scheduling strategy as a Markov decision process (MDP) and leveraged a DRL method called proximal policy optimization (PPO) to solve this problem. The proposed strategy learns the optimal offloading policy with the aim of minimizing the weighted long-term cost in terms of a tradeoff between task latency and energy consumption. Cui et al. \cite{rw_03} designed an IoV edge computing model and proposed an intelligent communication and computation resource allocation (ICCRA) for task offloading. They proposed a scheme combining multiple RL strategies, which are communication and computing resource allocation, to minimize the total system cost. 

Zhou et al. \cite{rw_04} formulated a joint optimization of offloading and resource allocation as a mixed-integer nonlinear programming (MINLP) problem in a multiuser MEC system with the objective of minimizing energy and using a double-deep-Q-network (DDQN) method as a solution. Waqar et al. \cite{rw_05} modeled an MEC-enabled integrated aerial-terrestrial vehicular network to minimize the overall computational overhead of the system. In this work, a joint computation offloading and resource allocation problem was formulated as a MINLP, and a multi-agent RL based on a Deep Double Q-learning (DDQN) algorithm was proposed to address this issue. Liu et al. \cite{rw_06} proposed a DRL technique aimed at determining optimal resource allocation and computation offloading policies. This approach takes into account the potential for mobile vehicles to offer computation services to user equipment (UE) within a vehicle edge computing network.  

Ning et al. \cite{rw_08} proposed a three-layer offloading framework in the IoVs modeled to minimize the overall energy consumption. Furthermore, their problem was divided into two parts which are the flow direction and offloading decision. They proposed the Edmonds-Karp algorithm to address the flow direction and DRL for the offloading decision. Wu et al. \cite{rw_09} proposed the Joint Computation Offloading and Task Migration Optimization (JCOTM) algorithm, based on the deep Q-network (DQN), to minimize the total system cost in a vehicle-aware Mobile Edge Computing Network (VAMECN). Xia et al. \cite{rw_10} proposed maximizing the capacity of the vehicular network to achieve efficient and reliable communication using cluster-enabled cooperative scheduling based on the Q-learning algorithm. Zhou et al. \cite{zhou2022} investigated the problem of jointly optimizing computation offloading, service caching, and resource allocation in collaborative MEC systems. They proposed a DDPG algorithm as a solution and formulated the problem as an MINLP, aiming to minimize long-term energy consumption. Budhiraja et al. \cite{budhiraja2022} proposed a latency-energy-aware task-offloading framework for connected autonomous vehicular (CAV) networks. The framework addresses the offloading of tasks and the reduction of energy consumption within the vehicle-to-infrastructure (V2I) and vehicle-to-vehicle (V2V) communication links of CAV networks. They formulated the problem as an MINLP and applied the soft actor-critic (SAC) based algorithm to jointly minimize latency and energy.

\section{System Model and Problem Formulation}

\begin{figure}[!t]
  \centering
  \includegraphics[width=8.5cm]{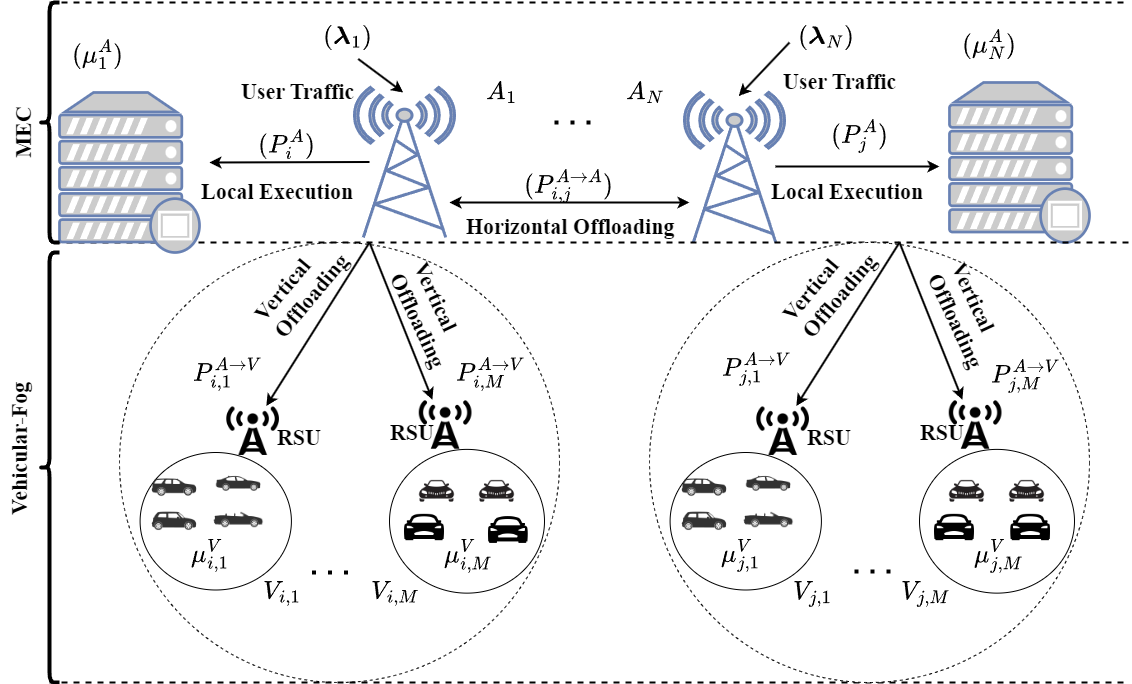}
  \caption{MEC and vehicular-fog architecture.}\label{fig:sysarch}
\end{figure}

In this paper, we consider a two-tier architecture comprising MEC and dynamic parking-lot VFs with vehicle arrivals and departures. The MEC sites are located behind base stations on the top tier (as depicted in Figure \ref{fig:sysarch}), while the VFs are positioned on the bottom tier. Table \ref{tab:notation} provides the notations used for modeling and problem formulation. Each MEC is denoted as ${A}_{i}$, where  $\text{\emph{i}}=\{1,2,\cdots,N\}$. We modeled each MEC ${A}_{i}$ in the topology as an ${M}\slash {M}\slash 1$ queueing system, following a first-come-first-serve (FCFS) manner. Each MEC ${A}_{i}$ is equipped with a single server having a huge computational capacity with a  mean service rate $\mu^{A}_{i}$ to process incoming traffic and a communication capacity $B^{A\rightarrow V}_{i,k}$ to communicate with $k^{th}$ VFs. Moreover, each MEC ${A}_{i}$ receives incoming hotspot traffic following a Poisson process with a mean arrival rate of $\lambda_{i}$.

All the VFs of the $i^{th}$ MEC site are denoted as $V_{i,k}$,  where $\text{\emph{k}}=\{1,2,\cdots,M\}$. The total number of vehicles in a single VF is denoted as $b_{k}$, where $b_{min} \leq b_{k} \leq b_{max}$. Here, $b_{min}$ is the minimum number of vehicles required to create a VF, and a VF will be suspended if $b_{k} < b_{min}$. Conversely, $b_{max}$ is the maximum number of vehicles allowed in a VF, and any arrivals will be prevented from entering the VF if $b_{k} \ge b_{max}$. We modeled each VF of the $i^{th}$ MEC site $V_{i,k}$ as $M\slash M\slash c$ queueing system with an FCFS manner, where each vehicle within the VF is considered as a single server with a mean service rate $\mu_{k,b}$, where $\text{\emph{b}}=\{1,2,\cdots,b_{k}\}$. Each VF $V_{i,k}$ of the $i^{th}$ MEC site has a computing capacity mean service rate $\mu^{V}_{i,k} = b_{k} \times \mu_{k,b}$ and a communication capacity $B^{V\rightarrow A}_{i,k}$. 

 In this paper, we consider both vertical federation, which involves the interaction between an MEC site and its associated VFs, and horizontal federation among different MEC sites.
 This implies that when incoming hotspot traffic arrives at MEC $A_{i}$, a portion of the traffic will be executed locally at the host MEC site $A_{i}$ with an offloading ratio of $P_{i}^{A}$. Another portion of traffic will be offloaded horizontally and executed at the neighbor MEC site $A_{j}$,   where $\forall A_{j} \in A_{N} \backslash \{A_{i}\}$, with an offloading ratio of $P_{i,j}^{A\rightarrow A}$. Additionally, some amount of traffic will be vertically offloaded and computed at the connected  $i^{th}$ MEC site VFs $V_{i,k}$ with an offloading ratio of $P_{i,k}^{A\rightarrow V}$. It is important to note that the sum of these diverse offloading ratios must equal 1.
\begin{table}[!t]
  \centering
  \caption{List of Commonly used Variables and Notations.}
    \begin{tabular}{|l|l|}
    \hline
    \multicolumn{1}{|c|}{\textbf{Notations}} & \multicolumn{1}{c|}{\textbf{Descriptions}} \\
    \hline
    $A_{i}$     & \multicolumn{1}{p{22.82em}|}{$i^{th}$ MEC site in tier-2, where $\text{\emph{i}}=\{1,2,\cdots,N\}$ } \\
    \hline
     $V_{i,k}$    & \multicolumn{1}{p{22.82em}|}{$k^{th}$ VF node in tier-1, where $\text{\emph{k}}=\{1,2,\cdots,M\}$} \\
    \hline
    $d^{A\rightarrow A}_{i,j}$     & \multicolumn{1}{p{22.82em}|}{Distance between $i^{th}$ MEC site and  $j^{th}$ MEC site} \\
    \hline
    $\mu^{A}_{i}$     & Computing Capacity of $i^{th}$ MEC site \\
    \hline
    $\mu^{V}_{i,k}$     & \multicolumn{1}{p{22.82em}|}{Computing Capacity of $k^{th}$ VF node of $i^{th}$ MEC site} \\
    \hline
    $B^{V\rightarrow A}_{i,k}, B^{A\rightarrow V}_{i,k}$     & \multicolumn{1}{p{22.82em}|}{Communication capacity of $k^{th}$ VF node of $i^{th}$ MEC site} \\
    \hline
      $\lambda_{i}$   & Arrival traffic at $i^{th}$ MEC site \\
    \hline
    $P^{A}_{i}$     & \multicolumn{1}{p{22.82em}|}{Arrival traffic being served at $i^{th}$ MEC site} \\
    \hline
    $P^{A\rightarrow A}_{i,j}$    & \multicolumn{1}{p{22.82em}|}{Arrival traffic at $i^{th}$ MEC site being  served at $j^{th}$ MEC site} \\
    \hline
    $P^{A\rightarrow V}_{i,k}$    & \multicolumn{1}{p{22.82em}|}{Arrival traffic being served at $k^{th}$ VF node of $i^{th}$ MEC site} \\
    \hline
    $L^{A}_{i}$    & Latency when served by $i^{th}$ MEC site \\
    \hline
    $L^{A\rightarrow A}_{i,k}$    & \multicolumn{1}{p{22.82em}|}{Latency when part of $i^{th}$ MEC sit traffic served by $j^{th}$ MEC site} \\
    \hline
    $L^{A\rightarrow V}_{i,k}$    & \multicolumn{1}{p{22.82em}|}{Latency when part of $i^{th}$ MEC site traffic served by $k^{th}$ VF node} \\
    \hline
    $L_{\text{sys}}$    & Average system latency \\
    \hline
    $D^{A\leftrightarrow A}_{i,k}$    & \multicolumn{1}{p{22.82em}|}{Propagation delay between $i^{th}$ and $j^{th}$ MEC site vice versa} \\
    \hline
    $E^{A\rightarrow V}_{i,k}$    & \multicolumn{1}{p{22.82em}|}{Energy consumed when served by $k^{th}$ VF node of $i^{th}$ MEC site} \\
    \hline
    $E_{i,total}$    & Energy consumed in $i^{th}$ MEC site \\
    \hline
    $E_{\text{sys}}$    & Average system energy consumption  \\
    \hline
    \end{tabular}%
  \label{tab:notation}%
\end{table}%

\subsection{Local Execution model}
For local execution, the traffic arriving at the MEC site $A_{i}$ comprises two types. The first type corresponds to a fraction of incoming user hotspot traffic, designated for local computation at MEC site $A_{i}$. The second type encompasses all potential incoming traffic from the neighboring MEC site $A_{j}$, intended for computation at MEC site $A_{i}$. 
So, the overall traffic load of a single MEC site $A_{i}$ will be the sum of these distinct types of arrival traffic, and it is denoted as $\lambda_{i}^{A} = P_{i}^{A} \times \lambda_{i} + \sum_{j} P_{j,i}^{A\rightarrow A} \times \lambda_{j}, \text{where }  \forall A_{j} \in A_{N} \backslash \{A_{i}\}$. The computational latency of the locally executed traffic can be calculated as
    \begin{flalign}
    L^{A}_{i} = \frac{\lambda^{A}_{i}}{\mu^{A}_{i}(\mu^{A}_{i} - \lambda^{A}_{i})} + \frac{1}{\mu^{A}_{i}},
    \end{flalign}
    \begin{flalign}
    L_{i, local} = P^{A}_{i} \times L^{A}_{i}.
    \end{flalign}
We neglected to compute the energy consumption of the $i^{th}$ MEC servers under the assumption that they are connected to the electrical network and have access to virtually unlimited energy.

\subsection{Horizontal Offloading Model}
Since we are considering a horizontal federation among the MEC sites, MEC site  $A_{i}$ can offload a portion of its traffic for computation at some of its neighbor MEC sites, such as $A_{j}$. This results in two forms of traffic that arrive at MEC site $A_{j}$. Firstly, a certain amount of incoming user hotspot traffic arrives at MEC site $A_{j}$, designated for local computation at that specific site. Secondly, all potential incoming traffic from neighboring MEC sites $A_{i}$ is offloaded and processed at MEC site $A_{j}$. The traffic load of a single MEC server $A_{j}$ is the summation of these different types of arrival traffic, denoted as $\lambda_{j}^{A} = P_{j}^{A} \times \lambda_{j} + \sum_{i} P_{i,j}^{A\rightarrow A} \times \lambda_{i}, \text{where } \forall A_{i} \in A_{N} \backslash \{A_{j}\}$. The computational latency of the horizontally offloaded and executed traffic can be calculated as
    \begin{flalign}
    L^{A \rightarrow A}_{i,j} = \frac{\lambda^{A}_{j}}{\mu^{A}_{j}(\mu^{A}_{j} - \lambda^{A}_{j})} + \frac{1}{\mu^{A}_{j}} + 2 \times D_{i,j}^{A\leftrightarrow A}.
    \end{flalign}
where $D_{i,j}^{A\leftrightarrow A} = \frac{d_{i,j}^{A\leftrightarrow A}}{c}$ represents the propagation delay between the $i^{th}$ and $j^{th}$ MEC sites, and $c$ is the speed of light. We assume that the host MEC site $A_{i}$ partially offloads its traffic to $h$ number of its neighbor MEC sites, denoted as $\text{\emph{h}} = \{1,2,\cdots,N\}$, where $h \leq N$ to prevent unnecessary delays that could occur by offloading to all $N$ neighbor MEC sites. The horizontal latency can be calculated by considering the maximum latency among the $h$ neighbor MEC sites, which are computed as the $i^{th}$ MEC traffic, and it is defined as
    \begin{flalign}
    L_{i, horiz} =\max(P^{A\rightarrow A}_{i,1} \times L^{A\rightarrow A}_{i,1},\cdots, P^{A\rightarrow A}_{i,h} \times L^{A\rightarrow A}_{i,h}).
    \end{flalign}
Here, we refrain from computing the energy consumption of the $j^{th}$ MEC site as we assume it is connected to the electrical network and has access to virtually unlimited energy.

\subsection{Vertical Offloading Model}
From the incoming user hotspot traffic $\lambda_{i}$ that arrives at the $i^{th}$ MEC site, some portions of it will be offloaded to be computed at the $k^{th}$ VF under the coverage of the same MEC site. The offloading ratio for this process is denoted by $P_{i,k}^{A\rightarrow V}$. The computational latency of vertically offloaded and executed traffic at $k^{th}$ VF is given in the form of Erlang’s C formula \cite{sm_01}.
    \begin{flalign}
    C\Bigl(b_{k}, \frac{\lambda^{V}_{i,k}}{\mu^{V}_{i,k}}\Bigr) = \frac{1}{1 + (1-\rho) \Bigl(\frac{b_{k}!}{(b_{k} \times \rho)^{b_{k}}}\Bigr) \sum_{k=0}^{b_{k}-1} \frac{(b_{k} \times \rho)^{k}}{k!}},
    \end{flalign}
where $\lambda_{i,k}^{V} = P_{i,k}^{A \rightarrow V} \times \lambda_{i}$ is the mean arrival rate of the $k^{th}$ VF, $\mu_{i,k}^{V}$ is the mean service rate of $k^{th}$ VF and $\rho=\frac{\lambda_{i,k}^{V}}{(b_{k} \times \mu_{i,k}^{V})}$  is the mean utilization. The transmission rate between the $k^{th}$ VF and $i^{th}$ MEC site is given as
    \begin{flalign}
    B^{A \rightarrow V}_{i,k} = W \times \log_{2}(1 + \frac{F_{i} \times G^{2}}{\omega \times W}),
    \end{flalign}
    \begin{flalign}
    B^{V \rightarrow A}_{i,k} = W \times \log_{2}(1 + \frac{F_{k} \times G^{2}}{\omega \times W}),
    \end{flalign}
where $B_{i,k}^{A \rightarrow V}$ is the downlink rate, $B_{i,k}^{V \rightarrow A}$ is uplink rate, $W$ is the uplink and downlink channel bandwidth, $F_{i}$  and $F_{k}$ are the transmission power of the $k^{th}$ VF and the $i^{th}$ MEC site, respectively. $\omega$ is the density of noise power, and $G$ is the channel gain between the $k^{th}$ VF and $i^{th}$ MEC site. The latency of the vertically offloaded traffic will be the sum of uplink, downlink, and computation delays, and it is defined as
    \begin{flalign}
    L^{A \rightarrow V}_{i,k} = \frac{\lambda^{V}_{i,k}}{B^{A\rightarrow V}_{i,k}} + \Biggl( \frac{C\Bigl(b_{k}, \frac{\lambda^{V}_{i,k}}{\mu^{V}_{i,k}}\Bigr)}{b_{k} \times \mu^{V}_{i,k} - \lambda^{V}_{i,k}} + \frac{1}{\mu^{V}_{i,k}} \Biggr) + \frac{\lambda^{V}_{i,k} \times \epsilon_{k}}{B^{V \rightarrow A}_{i,k}},
    \end{flalign}
where $\epsilon_{k}$ is the ratio of return traffic form the $k^{th}$ VF to the $i^{th}$ MEC site after completion of the execution. In this work, instead of offloading to all $M$ number of VFs, we only offload to $q$ number of VFs, which is denoted as $\text{\emph{q}} = \{1,2,\cdots, M\}$, where $q \leq M$. 
We have considered the maximum latency necessary for executing the offloaded traffic and is computed as
%
    \begin{flalign}
    L_{i, vertical} = \max(P^{A\rightarrow V}_{i,1} \times L^{A\rightarrow V}_{i,1},\cdots, P^{A\rightarrow V}_{i,q} \times L^{A\rightarrow V}_{i,q}).
    \end{flalign}

The energy consumption involves both the computation of all traffic arriving
at the $k^{th}$ VF and the data transfer between $i^{th}$ MEC site and $k^{th}$ VF, can be calculated as
   \begin{flalign} \label{eq:ecomp}
   E^{V}_{k,comp} = x_{k} \times y_{k} \times \lambda^{V}_{i,k} \times \vartheta,
   \end{flalign}
   \begin{flalign} \label{eq:edown}
    E^{A \rightarrow V}_{k,down} = F_{i} \frac{\lambda^{V}_{i,k} \times \vartheta}{B^{A \rightarrow V}_{i,k}},
   \end{flalign}
   \begin{flalign} \label{eq:eup}
    E^{V \rightarrow A}_{k,down} = F_{k} \frac{\lambda^{V}_{i,k} \times \vartheta \times \epsilon_{k}}{B^{V \rightarrow A}_{i,k}},
   \end{flalign}
where $E_{k,comp}^{V}$ is the $k^{th}$ VF computing energy, $E_{k,down}^{A\rightarrow V}$ is the downlink energy consumption, $E_{k,up}^{V\rightarrow A}$ is the uplink energy consumption, $x_{k}$ is the number of CPU cycles per bit, $y_{k}$ is the energy consumption per CPU cycle and $\vartheta$ is the packet size of the offloaded data. The total energy consumed at the $k^{th}$ VF to compute the incoming traffic, is represented by the sum of computing energy (Equation (\ref{eq:ecomp})), downlink energy (Equation (\ref{eq:edown})) and uplink energy (Equation (\ref{eq:eup})), as in Equation (\ref{eq:etotal}).
   \begin{flalign} \label{eq:etotal}
    E^{A \rightarrow V}_{i,k} = E^{A \rightarrow V}_{i,down} + E^{V}_{k,comp} + E^{V \rightarrow A}_{k,up}.
   \end{flalign}

\subsection{Problem Formulation}
The incoming user hotspot traffic that arrived at the $i^{th}$ MEC site is offloaded with some ratio and executed locally, horizontally, and vertically before returning back. The total latency of the task arrived at the $i^{th}$ MEC site, and its energy consumption is defined as
   \begin{flalign}
    L_{i} = \max(L_{i,local}, L_{i,horiz}, L_{i,vertical}),
   \end{flalign}
   \begin{flalign}
    E_{i,total} = \sum_{k}^{M} E^{A\rightarrow V}_{i,k}. 
   \end{flalign}
   The average system latency (Equation (\ref{eq:latency})) and average energy consumption (Equation (\ref{eq:energy})) will be calculated as
    \begin{flalign} \label{eq:latency}
        L_{\text{sys}} = \frac{1}{N} \sum^{N}_{i=1} L_{i},
    \end{flalign}
    \begin{flalign} \label{eq:energy}
        E_{\text{sys}} = \frac{1}{N} \sum^{N}_{i=1} E_{i, total}.
    \end{flalign}

Hence, we formulate the objective function with the aim of minimizing the weighted sum of system cost, considering both latency and energy. This objective function is defined as
    \begin{flalign}
    C_{\text{sys}} = \min(\sigma \times L_{\text{sys}} + (1-\sigma) \times E_{\text{sys}}),
    \end{flalign}
where $\sigma$ is the weighting parameter of execution latency and energy consumption, and the objective function is subjected to the following constraints: 
    \begin{flalign} \label{eq:const_1}
        \sigma + (1 - \sigma) = 1,
    \end{flalign}
    \begin{flalign} \label{eq:const_2}
        P^{A}_{i} + \sum_{h} P^{A\rightarrow A}_{i,h} + \sum_{q} P^{A\rightarrow V}_{i,q} = 1,
    \end{flalign}
    \begin{flalign} \label{eq:const_3}
        0 \leq  P^{A}_{i},  \{P^{A\rightarrow A}_{i,1} \cdots P^{A\rightarrow A}_{i,h}\},  \{P^{A\rightarrow V}_{i,1} \cdots P^{A\rightarrow V}_{i,q}\} \leq 1,
    \end{flalign}
    \begin{flalign} \label{eq:const_4}
     P^{A}_{i} \times \lambda^{A}_{i} + \sum_{h} P^{A\rightarrow A}_{i,h} \times \lambda^{A\rightarrow A}_{i,h} + \sum_{q} P^{A\rightarrow V}_{i,q} \times \lambda^{A\rightarrow V}_{i,q} \leq \lambda_{i},
    \end{flalign}
    \begin{flalign} \label{eq:const_5}
         \lambda_{i}^{A} \leq \mu^{A}_{i},
    \end{flalign}
    \begin{flalign} \label{eq:const_6}
         \lambda_{j}^{A} \leq \mu^{A}_{j},
    \end{flalign}
    \begin{flalign} \label{eq:const_7}
    \lambda_{i,k}^{V} \leq \mu^{V}_{i,k}.
    \end{flalign}
Equation (\ref{eq:const_1}) implies that the value of the weighted parameter must be between [0,1] and should sum to 1. Equation (\ref{eq:const_2}) and Equation (\ref{eq:const_3}) ensure that the offloading ratio must be between [0,1] and should add up to 1. Equation (\ref{eq:const_4}) ensures that the offloaded traffic doesn’t exceed the incoming hotspot traffic. Equation (\ref{eq:const_5}), Equation (\ref{eq:const_6}), and Equation (\ref{eq:const_7}) indicate that the offloaded traffic can be handled using the available resources in the network.

\section{Offloading with Distributed-TD3}
\begin{figure*}[!ht]
  \centering
  \includegraphics[width=16cm]{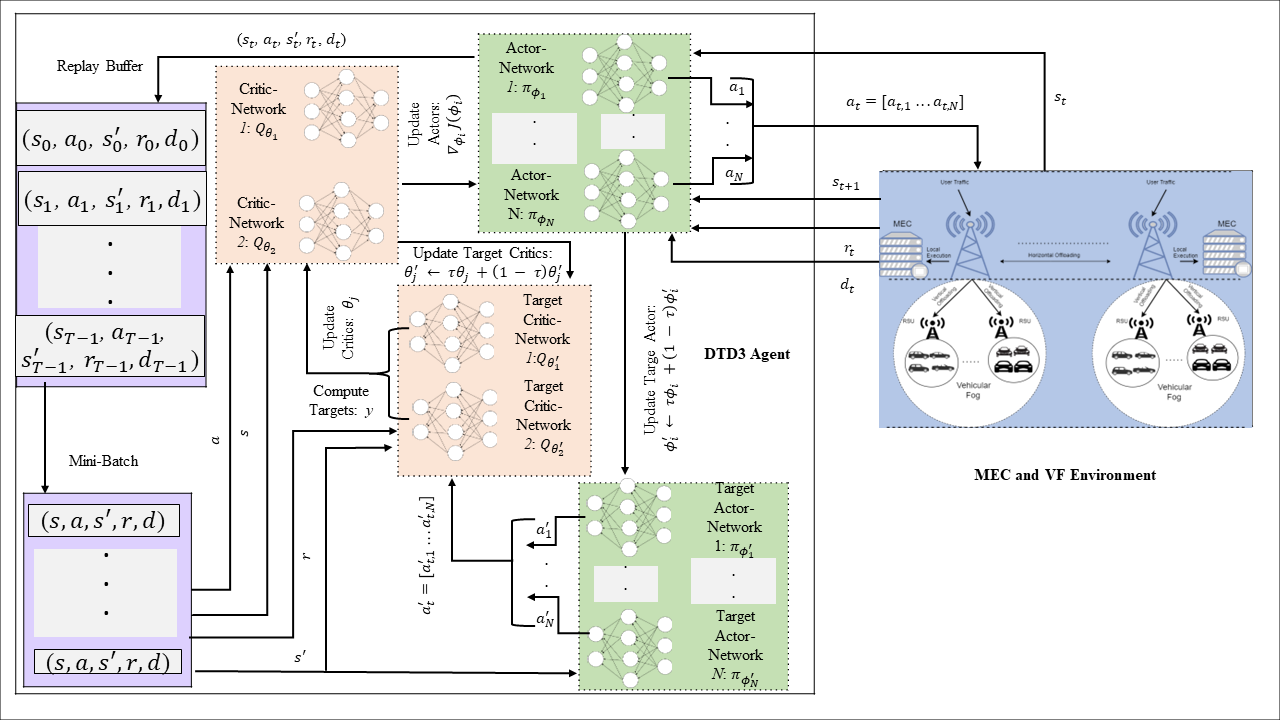}
  \caption{Distributed-TD3 model for MEC and VF architecture.}\label{fig:dtd3model}
\end{figure*}

The control plane in a two-tier MEC and VF system determines the best offloading decision to minimize system latency and energy consumption. RL is a potential method for offloading optimization, as it learns the optimal solution directly from the environment and can choose a sub-optimal solution without waiting for the best one. This offloading decision will be made using an RL agent. In RL strategy, an agent interacts with an unknown dynamic environment and takes different actions to maximize the total reward. At each discrete time step $t$, an agent observes a state  $s_{t}$ from a given state space $S$ and selects an action $a_{t}$ from an action space $A$ to transit from the current state $s_{t}$  to a new state $s_{t+1}$ following a policy $\pi({a_{t},s_{t}})$ and receives a reward $r_{t}$ based on the reward function $R (s_{t}, a_{t})$. This process is continued until the agent reaches the terminal state, where the main objective is to maximize the expected cumulative rewards \cite{ps_01}. 
In this paper, we transform the two-tier MEC and VF network into a MDP environment, and to determine offloading decisions in the given network topology, we use a Distributed-TD3 (DTD3) model as in Figure \ref{fig:dtd3model}, based on TD3 model \cite{ps_02}.

\subsection{Overview of TD3 Algorithm}
Twin Delayed Deep Deterministic Policy Gradients (TD3) \cite{ps_02} is a DRL algorithm designed for tasks with continuous action spaces. It builds on the Deep Deterministic Policy Gradients (DDPG) algorithm \cite{ddpg} to enhance stability, learning efficiency, and reduce overestimation by introducing the following key enhancements which make TD3 a valuable tool for training DRL agents in complex and continuous environments.

\emph{\textbf{1) Clipped Double-Q Learning:}} It is a technique used to address the problem of overestimation bias in value estimation. In Double-Q Learning, two value functions estimate the value of state-action pairs, and the minimum of these two estimates is used as the target value during updates. Clipping involves constraining the target value to be within a certain range, mitigating overestimation issues commonly observed in single Q-value estimators.

\emph{\textbf{2) Delayed Policy Updates:}} It is a strategy employed to improve training stability. Instead of updating the policy after every time step, updates are delayed. This delay allows the critic networks to provide more accurate and stable value estimates, reducing the potential for harmful policy oscillations and improving the overall efficiency of the learning process.

\emph{\textbf{3) Target Policy Smoothing:}} It is a technique used to stabilize the training of TD3 agents. It involves adding a small amount of noise to the target actions during the learning process. This smoothing of the target policy helps prevent the learning algorithm from becoming overly sensitive to small changes in the policy, contributing to more robust and conservative policy updates.

\subsection{MEC and VF Environment}
Offloading decisions in MEC and VF networks is a control task problem in which an appropriate decision must be made at each moment in time to achieve its objective. All control task problems can be represented in a general template called a MDP, which is a discrete-time stochastic control process. Therefore, the offloading decision in the MEC and VF network can be transformed into an MDP because it satisfies the MDP properties: it is a control process based on decision-making to achieve the objective, it is a stochastic process because the agent's actions partially affect the evolution of the offloading decision, and it's a discrete-time process because the offloading decision progresses in a finite interval of time. The details of the elements of the MDP for the MEC and VF network ($A_{i},V_{k}$) are presented as follows.

\emph{\textbf{1) State:}} The state space represents all the relevant information that describes the current condition of the MEC and VF environment at a given moment in time. We considered the following attributes to define the state space of the ME and VF environment: $\lambda={\{\lambda_{i}\}_{i=1}^{N}}$ which denotes a set of arrival hotspot traffic rates at all MEC sites; $\mu= \bigl\{ {\{\mu^{A}_{i}\}_{i=1}^{N}}, {\{\mu_{i,k}\}_{i=1, k=1}^{N, M}\bigr\}}$ is an array holding information about the computing capability at the MEC and VF; $B= \bigl\{B_{i,k}^{V\rightarrow A},B_{i,k}^{A\rightarrow V}\bigr\}_{i=1,k=1}^{N,M}$ denotes a set of communication capacities between MEC and vehicular-fog;  the average system latency $L_{sys}$; and the average system energy consumption $E_{sys}$. Thus, the state space $s_{t}$ at time $t$ is donated as $s_{t}=[\lambda, \mu, B, L_{sys}, E_{sys}]$ and can be observed by the agent to take and action.

\emph{\textbf{2) Action:}}  The action space produces an effect on the MEC and VF environment, leading to a modification of its state. Since the offloading decisions $(P_{i}^{A},P_{i,j}^{A\rightarrow A},P_{i,k}^{A\rightarrow V})$ are partial offloading decisions, the action space is continuous. Thus, the action space used in this environment is a combination of three different sets: $la_{t,i} = \{P_{i}^{A}\}_{i=1}^{N}$, which is denotes a set of all actions decided to be computed in $i^{th}$ MEC; $ha_{t,i}=\{P_{i,j}^{A\rightarrow A}\}_{j=1,}^{N},\forall {j} \in {N}\backslash{i}$ is a set of offloading decisions to be computed at neighbor MEC; and $va_{t,i}=\{P_{i,k}^{A\rightarrow V}\}_{i=1, }^{N}, \forall {k} \in {M} $ is a set of offloading decisions to be computed at connected VF. The combination of these three sets mentioned above will represent the $i^{th}$ MEC site action space $a_{t,i}$  at time $t$ as $a_{t,i}=[la_{t,i}, ha_{t,i},va_{t,i}]$, where $la_{t,i} + ha_{t,i} +va_{t,i} = 1$.

\emph{\textbf{3) Reward:}} Once an agent observes a state $s_{t}$ and takes an action $a_{t}$ as a result, the agent will observe the new state again and receive a reward from the MEC and VF environment. The reward is the feedback that an agent gets on the effect that the action $a_{t}$ had on the environment. In this work, the reward function is related to the objective function of our optimization problem $C_{\text{sys}}$. The reward $r_{t}$ is obtained by dividing the objective function $C_{\text{sys}}$ by some large number $\Omega$ to convert its value into fractions because neural networks work well with fractions rather than large decimal numbers. The reward value will be positive if the current value of the objective function $C_{\text{sys}}$ is less than or equal to the previous value of the objective function $C_{\text{sys}}'$, and it will be negative if otherwise. The reward is defined as,

\begin{equation} \label{eq:ps_00}
r_t=
    \begin{cases}
    \frac{C_{\text{sys}}}{\Omega}, & C_{\text{sys}} \leq C_{\text{sys}}',\\ 
     \frac{-C_{\text{sys}}}{\Omega}, & Otherwise.
     \end{cases}
\end{equation}
    
\emph{\textbf{4) Done:}} The done element is another important aspect of the MEC and VF environment that informs the agent whether it has reached the terminal state or not. In this environment, there are two ways the agent will know if it is in the terminal state or not. The first one is after ten different offloading decisions, the environment will reach the terminal state, and the second one is if,  at any point during the simulation, the agent provides an offloading decision that puts any of the MEC or VF resources into a busy state, the environment will terminate automatically.

\subsection{Proposed Distributed-TD3 Algorithm}
In this paper, we propose the DTD3 algorithm (Algorithm \ref{alg:DTD3}), which builds on TD3. The DTD3 agent is deployed in the control plane and is responsible for finding an optimal offloading destination by taking actions (offloading decisions) that maximize its expected reward.

In the MEC and VF environment, there are $N$ MEC sites $A_{1},A_{2},\cdots,A_{N}$ that receive an incoming hotspot traffic $\lambda_{1},\lambda_{2},\cdots,\lambda_{N}$ from users. Hence, we need $N$ different actions $a_{1},a_{2},\cdots,a_{N}$, where each action sums up to 1, and they are merged together to form the action at time $t$, $a_{t}=[a_{t,1}, a_{t,2}, \cdots, a_{t,N}]$. To achieve this, we use a distributed policy that enables us to take action for each MEC site independently and then concatenate them together to send to the environment to have an impact. 

For this purpose, we employ $N$ actor policy networks $\pi_{\phi_{1} },\cdots,\pi_{\phi_{N}}$ with parameters $\phi_{1},\cdots,\phi_{N}$ and actor-target policy networks $\pi_{\phi_{1}'},\cdots,\pi_{\phi_{N}'}$ with parameters $\phi_{1}',\cdots,\phi_{N}'$ that work in parallel to find an optimal offloading decision and merge their results at the end. Additionally, we use two critic networks $Q_{\theta_{1}}, Q_{\theta_{2}}$ with parameters $\theta_{1},\theta_{2}$ and critic-target networks $Q_{\theta_{1}'}, Q_{\theta_{2}'}$ with parameters $\theta_{1}', \theta_{2}'$ to minimize and correct overestimation.

To optimize our neural networks, we use a replay buffer $\mathcal{B}$ to store and sample experiences. At each time $t$, a DTD3 agent observes the state $s_{t}$ and takes an action $a_{t}$ using the actor policy networks (Equation (\ref{eq:ps_01})) and (Equation (\ref{eq:ps_02})).
    \begin{flalign} \label{eq:ps_01}
        a_{i}(s_{t}) = clip (\pi_{\phi_{i}}(s_{t}) + \epsilon , a_{min},a_{max}), \nonumber \\
        \epsilon \sim N (0,\sigma) \hspace{1em} for \hspace{1em} i=1,\cdots,N.
    \end{flalign}
    \begin{flalign} \label{eq:ps_02}
        a_{t}(s_{t}) =[a_{t,1}(s_{t}), a_{t,2}(s_{t}), \cdots, a_{t,N}(s_{t})].
    \end{flalign}
where $\epsilon$ is Gaussian noise, and $a_{min}$ and $a_{max}$ are the lower and higher values of the action space that an agent can choose from. After the action is executed in MEC and VF environment, the agent will receive the next state $s_{t}'$, reward $r_{t}$, and done flag $d_{t}$ alongside the previous state $s_{t}$ and action $a_{t}$ as a single transition $(s_{t}, a_{t}, s'_{t}, r_{t}, d_{t})$, which is store in replay buffer $\mathcal{B}$.

\begin{algorithm}[!t]
\small
\caption{Proposed Distributed-TD3 Algorithm} 
\label{alg:DTD3}
        \LinesNumbered
        \textbf{Input}:  Arrival hotspot traffic and No. of MEC $N$ \\
        \textbf{Output}:  Offloading ratio \\
        \textbf{1. Initialization:}\\
        Initialize actor networks $\pi_{(\phi_{1} )},\cdots,\pi_{(\phi_{N})}$ with random parameters $\phi_{1},\cdots,\phi_{N}$; \\
        Initialize target actor networks $\pi_{(\phi'_{1})},\cdots,\pi_{(\phi'_{N})}$ with random parameters $\phi'_{1},\cdots,\phi'_{N}$; \\
        Initialize critic networks $Q_{(\theta_{1})}$, $Q_{(\theta_{2})}$ with random parameters $\theta_{1}$, $\theta_{2}$;  \\
        Initialize target critic networks $Q_{(\theta'_{1})}$, $Q_{(\theta'_{2})}$ with random parameters $\theta'_{1}$, $\theta'_{2}$;\\
        Initialize replay buffer $\mathcal{B}$; \\
        \For{$t=1$ to $T$}
        {
            \textbf{2. Exploration and Generate training data:} 
            Observe state $s_{t}$\;
            Select actions 
            $a_{i}(s_{t}) = clip (\pi_{\phi_{i}}(s_{t}) + \epsilon , a_{min},a_{max}), \epsilon \sim N (0,\sigma)$ for $i=1,\cdots,N$\;
            $a_{t}(s_{t}) =\left[ a_{t,1} (s_{t}),a_{t,2}(s_{t}),\cdots,a_{t,N}(s_{t}) \right]$\;
            Observe next state $s'_{t}$ reward $r_{t}$, and done flag $d_{t}$ to indicate $s'_{t}$ is the terminal state save transition $(s_{t},a_{t},s'_{t},r_{t},d_{t})$ in replay buffer $\mathcal{B}$\; 
            \textbf{3. Learning or Exploitation:}
	      Sample random mini-batch of $M$ transitions 
            $(s, a, s', r, d)$ from $\mathcal{B}$\;
	      \If{$d = 0$}
            {reset environment state;}
            Select target actions:
            $a'_{i}(s') = clip (\pi_{\phi'_{i}}(s') + clip(\epsilon, -c, c), a_{min}, a_{max}), \epsilon \sim N (0,\sigma)$ for $i=1,\cdots,N$\;
            $a'_{t}(s') =\left[ a'_{t,1} (s'),a'_{t,2}(s'),\cdots,a'_{t,N}(s') \right]$\;
            Compute target value  $y = r + \gamma \min_{\theta_{j=1,2}} Q_{\theta'_{j}}(s', a'_{t})$\;
            Update critic networks \\ 
            $\theta_{j} \leftarrow argmin_{\theta_{j=1,2}} \frac{1}{M} \sum \left( y - Q_{\theta_{j}}(s,a)\right)^2$\;
            \If{$t$ \text{mod} $t'=0$}
            { 
                   Update actor networks using deterministic policy gradient
                   $\nabla_{\phi_{i}} J(\phi_{i}) = \frac{1}{M}\sum\nabla_{a} Q_{\theta_{1}}(s,a)\mid a $ \\ $= \pi_{\phi_{i}}(s) \nabla_{\phi_{i}} \pi_{\phi_{i}}(s)$, 
                   for $i=1,\cdots,N$\;
                   Update target actor networks:\\
                   $\phi'_{i} \leftarrow \tau \phi_{i} + (1-\tau) \phi'_{i}$,
                   for $i=1, \cdots, N$\;
                   Update target critic networks:\\
                   $\theta'_{j=1,2} \leftarrow \tau \theta_{j} + (1-\tau) \theta'_{j}$\;        
	}	
 }   
\end{algorithm}

Once we have enough experience in the buffer, we update the neural network by taking a random mini-batch of $M$ transitions $(s, a,s',r,d)$ from the buffer. To compute the target value $y$ in Equation (\ref{eq:ps_05}), we select the actions $a_t'$ that can be taken in the next state $s'$ using the actor-target policy. Subsequently, we apply clip noise to \textbf{smooth the target policy} and ensure it never gets stuck selecting an overestimating action as in Equations (\ref{eq:ps_03}) and (\ref{eq:ps_04}).
%
    \begin{flalign} \label{eq:ps_03}
        a'_{i}(s') = clip (\pi_{\phi_{i}'}(s') + clip(\epsilon, -c,c) , a_{min},a_{max}), \nonumber \\
        \epsilon \sim N (0,\sigma) \hspace{1em} for \hspace{1em} i=1,\cdots,N.
    \end{flalign}
    \begin{flalign} \label{eq:ps_04}
        a_{t}'(s') =[a'_{t,1} (s'),a_{t,2}'(s'),\cdots,a_{t,N}'(s')].
    \end{flalign}
We use both critic-target networks to estimate the Q-values for the next state-action pairs, employing \textbf{clipping} to prevent overestimation by selecting the value with the minimal estimate. Once we have the target value $y$ as in Equation (\ref{eq:ps_05}), we update each of these critic networks $\theta_{j}$ by calculating the mean squared error of these batch of observations and applying a step of gradient descent (Equation (\ref{eq:ps_06})).
    \begin{flalign} \label{eq:ps_05}
        y = r + \gamma \min_{\theta_{j=1,2}} Q_{\theta_{j}'}(s',a'_{t}). 
    \end{flalign}
    \begin{flalign} \label{eq:ps_06}
        \theta_{j} \leftarrow argmin_{\theta_{j=1,2}} \frac{1}{M} \sum (y - Q_{\theta_{j}}(s,a))^2.
    \end{flalign}
%
%
At every $t'$ time step, which corresponds to a \textbf{delayed policy update} interval, we update the actor policy $\phi_{i}$ by estimating the Q-value of the state and the action selected by the policy using one of the critic networks. 
Subsequently, we implement a gradient ascent step to adjust the parameters of the neural network $\phi_{i}$ in the direction of maximizing the Q-value (Equation (\ref{eq:ps_07})).
After updating the neural networks for the actor policy, we proceed to update the target networks. 
For this, we use Equation (\ref{eq:ps_08}) and Equation (\ref{eq:ps_09}) to update the target actor networks $\phi'_{i}$ and target critic networks $\theta'_{j}$, respectively, to achieve more stable and effective learning.
%
    \begin{flalign} \label{eq:ps_07}
        \nabla_{\phi_{i}} J(\phi_{i}) = \frac{1}{M}\sum\nabla_{a} Q_{\theta_{1}}(s,a)|a = \pi_{\phi_{i}}(s) \nabla_{\phi_{i}} \pi_{\phi_{i}}(s),
        \nonumber \\ for \hspace{1em} i=1,\cdots,N.
    \end{flalign}
    \begin{flalign} \label{eq:ps_08}
        \phi_{i}'  \leftarrow \tau \phi_{i}  + (1- \tau)  \phi_{i}' \hspace{1em}  for \hspace{1em} i=1,\cdots,N.
    \end{flalign}
    \begin{flalign} \label{eq:ps_09}
        \theta_{j=1,2}'  \leftarrow \tau \theta_{j}  + (1- \tau)  \theta_{j}'.
    \end{flalign}
This process will continuously repeat until the training epoch $t$ reaches $T$. Along the way, the agent will be able to provide the optimal offloading decisions that satisfy the requirements of the objective function in the given MEC and VF environment. 

\section{Results and Discussions}
In this section, we provide a comprehensive overview of our experimental setup for the simulation and conduct an in-depth analysis of the obtained results.
\subsection{Simulation Parameter Settings}
In this simulation, we have created an MEC and VF network environment along with its DTD3 agent, which is deployed on the control plane to determine the optimal offloading decision. The MEC and VF networks consist of four MEC sites, each connected to five VFs. Each VF represents a group of 5 to 25 parked vehicles. To ensure a realistic layout, the MEC sites are evenly distributed between distances of 1 to 10 km, utilizing the random geographic coordinate sampling application \cite{ra_01}. The computing capacity of each MEC site is set to 30 MIPS, while the computational capacity of the VF varies based on the number of vehicles, ranging from 3 to 15 MIPS. The transmission and reception power of the MEC and VF networks are set to 24 dBm, and the system bandwidth is 50 MHz. Incoming traffic from users arrives at each MEC site as either normal traffic (with rates of 10M, 20M, and 30M packets per second) or hotspot traffic (with rates of 40M, 60M, and 80M packets per second). For simplicity, we consider one million instructions per second (MIPS) and one million packets per second (MPPS) as equivalent. We developed the MEC and VF environment following the standards set by OpenAI Gym \cite{Greg2016}. Through rigorous testing, we assessed the environment's effectiveness and ensured its satisfaction with all necessary criteria for an RL environment, utilizing Stable-Baselines3 \cite{ra_02}. This enabled us to conduct a comprehensive assessment of the system's performance and confirm its suitability for comparing various DRL algorithms.

\begin{table}[!t]
  \centering
  \caption{Parameter settings}
    \begin{tabular}{|l|l|}
    \hline
    \multicolumn{1}{|c|}{\textbf{Parameters }} & \multicolumn{1}{c|}{\textbf{Value }} \\
    \hline
    Number of MEC $(A_{i})$   & 4 \\
    \hline
    Number of VF node of per MEC $(V_{i,k})$   & 5 \\
    \hline
    Number of vehicles per VF $(b_{k})$  & 5 to 25 \\
    \hline
    Distance between MECs $(d^{A\rightarrow A}_{i,j})$   &  1km to 10km \\
    \hline
    System Bandwidth $(W)$ & 50 MHz \\
    \hline
    Background noise $(\omega)$   & -110dBm \\
    \hline
    VF transmission and reception power $(F_{i}, F_{k})$  & 24dBm \\
    \hline
    The capacity of MEC server $(\mu^{A}_{i})$   & 30 MIPS \\
    \hline
    The capacity of $k^{th}$ VF per MEC site $(\mu^{V}_{i,k})$  & 3 to 15 MIPS \\
    \hline
    CPU cycles per bit $(x_{k})$  & 1900 cycles/bit \\
    \hline
    Energy consumption per CPU cycle $(y_{k})$  & 0.1 joule/cycle \\
    \hline
    Normal traffic rate Poisson-Dist $(\lambda_{i})$   & [10M, 20M, 30M p\slash s] \\
    \hline
    Hotspot traffic rate Poisson-Dist $(\lambda_{i})$  & [40M, 60M, 80M p\slash s] \\
    \hline
    Learning rate $(\alpha)$  & 0.0003 \\
    \hline
    Update rate $(\tau)$    & 0.005 \\
    \hline
    Discount factor $(\gamma)$    & 0.99 \\
    \hline
    Batch Size $(M)$ & 100 \\
    \hline
    Number of hidden layers & 2 \\
    \hline
    Number of neurons in hidden layer  & 256 \\
    \hline
    \end{tabular}%
  \label{tab:ps}%
\end{table}%



To evaluate the performance of the proposed DTD3 algorithm, we conducted a comparison with four other algorithms.
These include two from a DRL-based method, namely TD3 \cite{ps_02} and DDPG \cite{ddpg}, and two from traditional optimization approaches, Simulated Annealing (SA) \cite{ra_04} and Particle Swarm Optimization (PSO) \cite{salehizadeh2009local}. 
TD3 extends the original DDPG algorithm to address specific challenges and enhance stability during training. On the other hand, SA and PSO emerge as the preferred solutions for optimizing the offloading problem due to their adeptness in navigating local optima and exploring various regions within the solution space. 
%
%
Given the significance of hyperparameters in RL and their impact on the learning process, selecting appropriate values is crucial. To address this, we employed Optuna \cite{ra_03}, an automatic hyperparameter optimization software framework, to determine the right hyperparameter values for our simulation. 
In the DTD3 model, several actors and target networks were employed, contrasting with the TD3 model that utilized a single actor and target network for action selection. Both algorithms incorporated two critics and target networks, dedicated to evaluating the actor and its respective target networks.
The number of hidden layers and neurons in each network was kept the same. For the simulation, we set the episode size, memory size, and mini-batch size to 10000, 100000, and 100, respectively. The learning rate, update rate, discount factor, number of hidden layers, and number of neurons in each hidden layer of the actor and critic networks were all set to 0.0003, 0.005, 0.99, 2, and 256, respectively. To implement the simulation MEC and VF environments, we utilized Python, Gym, Pytorch, and NumPy. The essential parameter settings are summarized in Table \ref{tab:ps}.

\begin{figure}[!t]
  \centering
  \includegraphics[width=8.5cm]{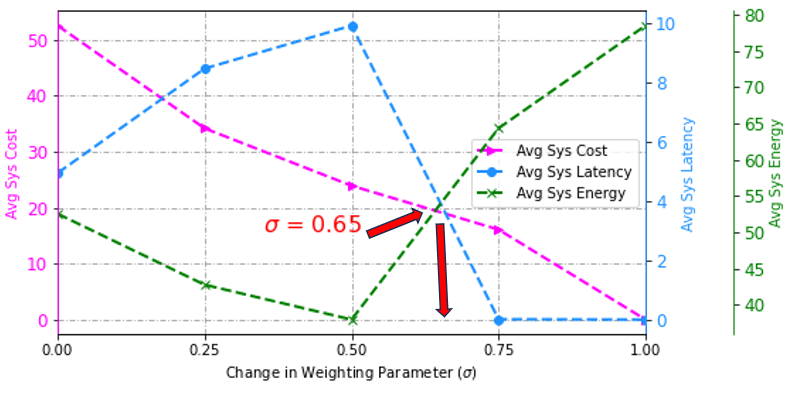}
   \caption{Performance with change in weighting parameter ($\sigma$).}\label{fig:wp}
\end{figure}

\begin{figure*}[!ht]
  \centering
  \subfloat[DTD3 model episode reward.]{
  \includegraphics[width=0.28\textwidth]{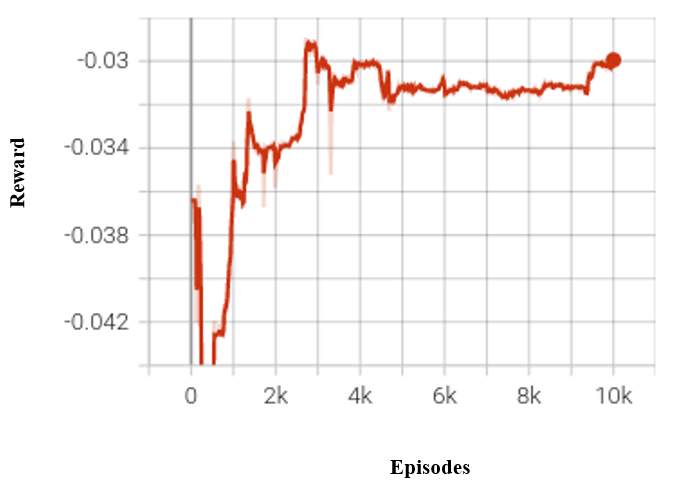}
   \label{fig:ep_reward}}
   \qquad
   \subfloat[DTD3 model maximum episode reward.]{
    \includegraphics[width=0.28\textwidth]{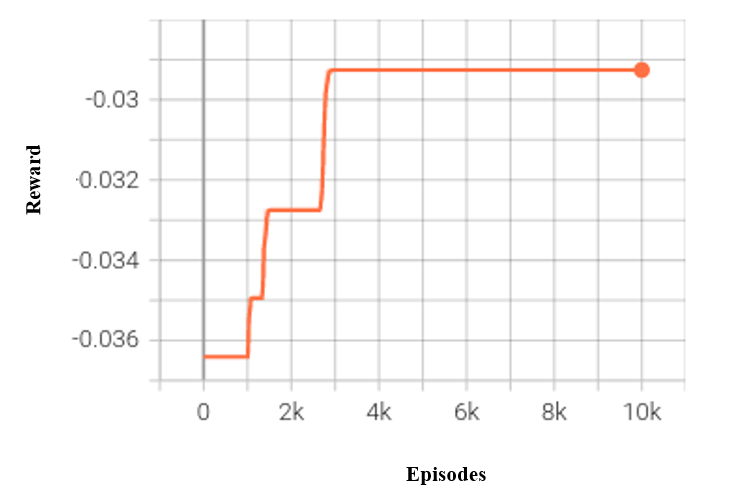}
   \label{fig:max_reward}}
   \qquad
   \subfloat[DTD3 model mean episode reward.]{
    \includegraphics[width=0.28\textwidth]{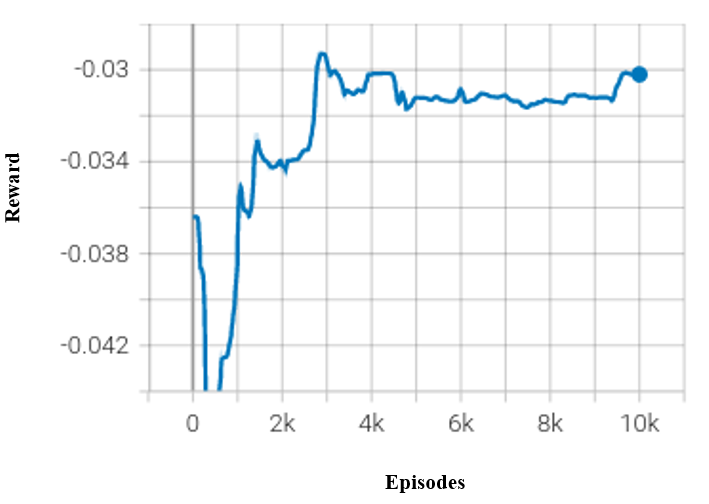}
   \label{fig:mean_reward}}

  \caption{DTD3 model performance in a single run.}
  \label{fig:reward}
\end{figure*}

Figure \ref{fig:wp}, illustrates the relationship between the average system cost, average system latency, and average system energy. To determine the best-weighted parameter value for the simulation, we conducted an experiment using several weighting parameter values (0, 0.25, 0.5, 0.75, and 1)  plotted on the x-axis. The figure uses three color-coded y-axis bars to represent the three performance attributes: the left bar in purple represents the average system cost, the middle bar in light blue represents the average system latency, and the right bar in green represents the average system energy. The graph shows that when the weighting parameter value is between 0 and 0.5, the average system cost and energy values decrease, while the average system latency value rises. This indicates that the average system cost value is more influenced by the average system energy value. On the other hand, when the weighting parameter value is between 0.5 and 1, the average system cost and latency values decrease, and the average system energy value increases, indicating that the average system cost value is more influenced by the average system latency value than the average system energy value. In this experiment, we set the weighting parameter ($\sigma$) to 0.65, as indicated by a red arrow in Figure~\ref{fig:wp}. The decision is based on the fact that, at this juncture, both the average system latency and energy make equal contributions to the overall system cost.


\subsection{Performance Analysis}
In this subsection, we evaluated the performance of the algorithms in terms of convergence, arrival traffic rate, and number of vehicles to form VFs.


\subsubsection{DTD3 Agent Performance Evaluation}

We present the learning progress of a DTD3 algorithm in a single-run experiment, as illustrated in Figure \ref{fig:reward}. The main objective of the agent is to maximize the long-term cumulative reward, achieved through the reward function defined in Equation (\ref{eq:ps_00}). Specifically, Figure \ref{fig:ep_reward} displays the episode reward, Figure \ref{fig:max_reward} shows the maximum reward achieved, and Figure \ref{fig:mean_reward} illustrates the average episode reward of the agent over the previous 100 runs during the learning process. In general, this figure shows how the agent has been learning progressively by receiving higher and higher rewards over time. This learning trend demonstrates the effectiveness of the DTD3 algorithm in optimizing the offloading decision-making process and achieving better overall performance as the number of episodes increases.

\begin{figure}[!t]
  \centering
  \includegraphics[width=7cm]{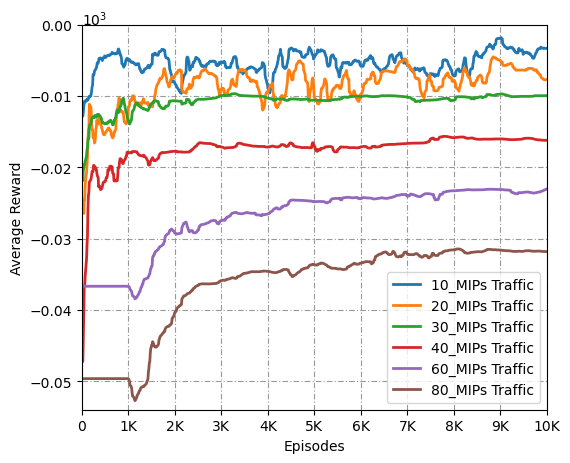}
   \caption{DTD3 model performance with different arrival traffic.}\label{fig:reward_traffic}
\end{figure}

\begin{figure}[!t]
  \centering
  \includegraphics[width=8cm]{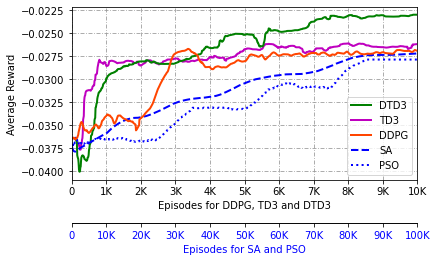}
   \caption{The convergence performance of DTD3, TD3, DDPG, SA, and PSO algorithms.}\label{fig:dtd3_sa_converg}
\end{figure}

\begin{figure*}[!t]
 \centering
  \subfloat[Average system cost.]{
  \includegraphics[width=0.35\textwidth]{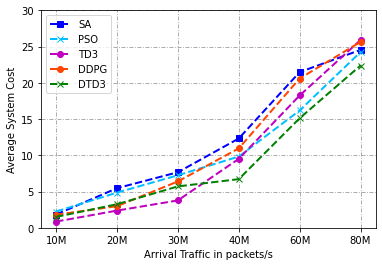}
   \label{fig:cost_traffic}}
    \qquad
  \subfloat[Average system energy consumption.]{
  \includegraphics[width=0.35\textwidth]{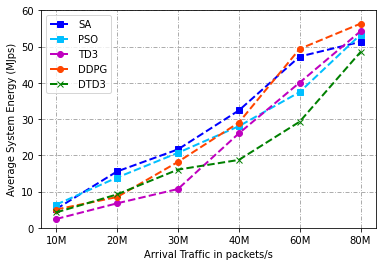}
   \label{fig:energy_traffic}}
    \qquad
  \subfloat[Average system latency.]{
  \includegraphics[width=0.35\textwidth]{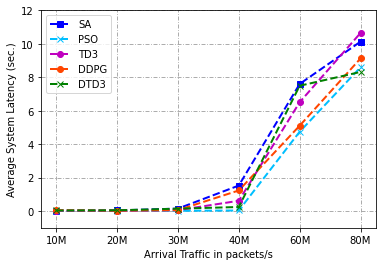}
   \label{fig:latency_traffic}}
    \qquad
  \subfloat[Average system utilization.]{
  \includegraphics[width=0.35\textwidth]{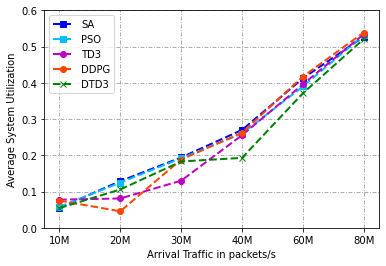}
   \label{fig:utilization_traffic}}
    \caption{The evaluations of system performance with change in traffic.}
    \label{fig:traffic}
\end{figure*}

In Figure \ref{fig:reward_traffic}, we consider running ten experiments and averaging the results due to the randomness in the MEC and VF networks to showcase the performance of the DTD3 model.
It illustrates the average reward with respect to the episodes for different arrival traffic rates. We can see that the agent is learning over time for normal traffic at 10M and 20M packets/s; however, the result is unstable and cannot converge. As the arrival traffic exceeds 30M packets/s, the agent learns very well by receiving higher average rewards throughout training, and after it reaches around 2K episodes, the difference in reward value becomes lower. As a result, the DTD3 agent converges to its optimal value for arrival hotspot traffic after 8K episodes.


In Figure \ref{fig:dtd3_sa_converg}, we present the convergence curves of the above-mentioned five algorithms. For the DRL-based approaches (DTD3, TD3, and DDPG), the total number of episodes used in the experiment is 10K, whereas the total number of episodes for the traditional approaches (SA and PSO) is 100K. Initially, both DTD3 and TD3 performed poorly in the first 100 episodes, receiving lower average rewards, and DDPG also performed poorly in the first 2K episodes. However, after this initial phase, all three algorithms start to learn successfully and receive higher rewards as the number of episodes increases. Between 1K to 2K episodes, TD3 begins to achieve significantly higher rewards compared to both DTD3 and DDPG. Then, between 2K to 3K episodes, both TD3 and DTD3 perform similarly. Between 3K to around 4K episodes, DDPG begins to achieve significantly higher rewards compared to both TD3 and DTD3. However, after 4K episodes, DTD3 surpasses both TD3 and DDPG in terms of rewards, and all three eventually stop growing and converge to their optimal values around 8K episodes. As the number of episodes increases, both the SA and PSO algorithm's rewards also improve and converge to their optimal value after 90K episodes. However, PSO performs less effectively compared to SA because of its tendency to settle in suboptimal solutions. As depicted in Figure~\ref{fig:dtd3_sa_converg}, all approaches improve their rewards and converge to their optimal values by making optimal offloading decisions. Notably, the DRL-based algorithms converge approximately ten times faster than the traditional optimization algorithms. Furthermore, DTD3 outperforms TD3 and DDPG, converging slightly faster.

\subsubsection{Performance Analysis with Change in Arrival Traffic}

Figure \ref{fig:traffic} shows the performance of the algorithms as the number of arrival traffic rates increases in terms of average system cost, energy consumption, latency, and utilization, as shown in Figures \ref{fig:cost_traffic}, \ref{fig:energy_traffic}, \ref{fig:latency_traffic}, and \ref{fig:utilization_traffic}, respectively. 

Figure \ref{fig:cost_traffic} shows that as the number of arrival traffic rates increases, the average system cost increases for all algorithms. In the scenario of normal traffic, TD3 outperformed other algorithms including DTD3. Additionally, both DTD3 and DDPG surpassed SA and PSO. However, when considering hotspot traffic, DTD3 emerged as the top performer among all approaches.
This is because TD3 and DTD3 can make more effective offloading decisions for incoming normal and hotspot traffic, respectively, to achieve a minimum average system cost.

Figure \ref{fig:energy_traffic} shows that TD3 achieves lower average system energy consumption when handling normal traffic compared to DTD3, DDPG, SA, and PSO. 
However, when the arrival traffic exceeds 30M packets/s, the energy consumption in TD3 continues to increase, while that of DTD3 becomes the lowest.
Specifically, when the arrival traffic reaches 80M packets/s or higher, TD3, DDPG, and PSO consume more average system energy than SA and DTD3 remains the most efficient among all algorithms. 
As shown in Figure \ref{fig:energy_traffic}, in the case of normal traffic, DTD3 and DDPG consume slightly more energy compared to TD3, however, it is relatively less than SA and PSO. Conversely, in the case of hotspot traffic, the energy consumption in DTD3 continues to be the least among all algorithms.


Figure \ref{fig:latency_traffic} shows that the average system latency increases for all algorithms as the number of arriving traffic packets increases. Initially, the latency appears similar across all algorithms until the arrival traffic reaches 30M packets/s. However, when the arrival traffic exceeds 30M packets/s, DTD3, TD3, and PSO outperform both DDPG and SA. Specifically, when the hotspot traffic exceeds 60M packets/s, DTD3 executes the traffic with a lower average system latency than both SA and TD3. Additionally, as the traffic approaches 80M packets/s, DTD3 outperforms both DDPG and PSO.

Figure \ref{fig:utilization_traffic} shows that SA and PSO utilize more computational resources than DRL-based methods for computing normal traffic. Conversely, for processing hotspot traffic, DDPG, SA, and PSO consume more resources. This is because all algorithms offload most of the traffic to the VFs rather than the MECs to utilize the distributed resource. 
While TD3, DDPG, SA, and PSO effectively manage normal traffic, as the arrival traffic increases, the VFs also become overloaded. This leads to higher average system energy consumption and system latency when handling hotspot traffic, resulting in a higher average system cost.
However, DTD3 avoids this issue by offering better offloading decisions, which yield comparable performance in normal traffic scenarios and superior performance in handling hotspot traffic compared to others across all cases.

\begin{figure*}[!t]
 \centering
    \subfloat[Average system cost.]{
        \includegraphics[width=0.45\textwidth]{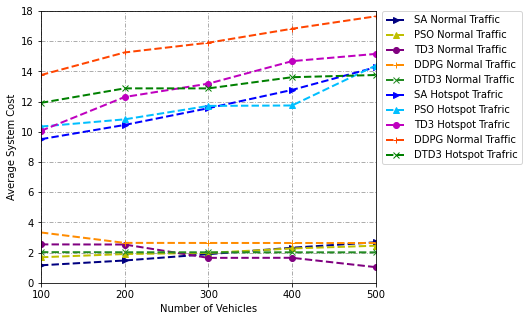}
        \label{fig:cost_vehicle}}
    \qquad
    \subfloat[Average system energy consumption.]{
        \includegraphics[width=0.45\textwidth]{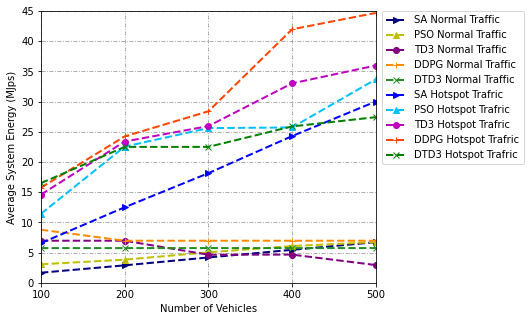}
        \label{fig:energy_vehicle}}
    \qquad
    \subfloat[Average system latency.]{
        \includegraphics[width=0.45\textwidth]{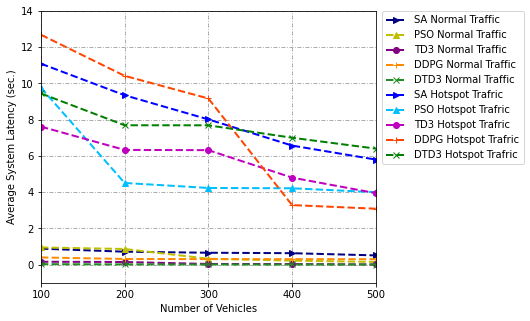}
        \label{fig:latency_vehicle}}
    \qquad
    \subfloat[Average system utilization.]{
        \includegraphics[width=0.45\textwidth]{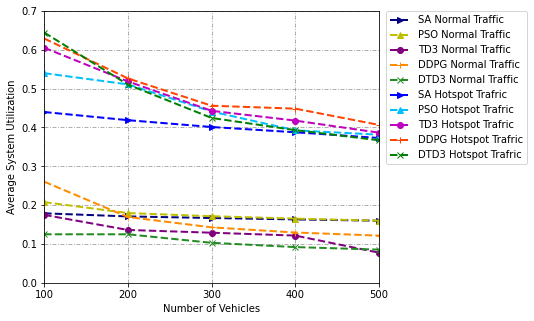}
        \label{fig:utilization_vehicle}}
    \caption{ The evaluations of system performance with changes in the number of vehicles.}
    \label{fig:vehicle}
\end{figure*}

\subsubsection{Performance Analysis with Change in Number of Vehicles in the VFs}
Figure \ref{fig:vehicle} illustrates the performance of algorithms as the number of vehicles in the VFs increases in terms of average system cost, energy consumption, latency, and utilization, as shown in Figures \ref{fig:cost_vehicle}, \ref{fig:energy_vehicle}, \ref{fig:latency_vehicle}, and \ref{fig:utilization_vehicle}, respectively. 

Figure \ref{fig:cost_vehicle} shows the relationship between the number of vehicles in the VF and the average system cost for both normal and hotspot traffic. When the number of vehicles in the VF is less than 300, traditional methods outperform the DRL-based methods. 
Conversely, when the number of vehicles exceeds 300, both TD3 and DTD3 perform better than DDPG, SA, and PSO. For hotspot traffic, DTD3 surpasses TD3 when the number of vehicles in the VF exceeds 300. Moreover, when the number of vehicles exceeds 400, DTD3 outperforms both SA and PSO.

Figure \ref{fig:energy_vehicle} shows the relationship between the number of vehicles in VF and average system energy consumption for both normal and hotspot traffic. SA and PSO consume less average system energy than DRL-based methods when executing normal traffic until the number of vehicles reaches 400. However, once the number of vehicles exceeds 400, both TD3 and DTD3 consume less average system energy than DDPG, SA, and PSO. The energy consumption by SA for computing hotspot traffic increases linearly. In contrast, DTD3 initially performs worse than PSO until the number of vehicles exceeds 200 and SA until the number of vehicles in the VF exceeds 400, at which point it starts consuming less energy. However, both TD3 and DDPG perform less effectively compared to SA, PSO, and DTD3.

In Figure \ref{fig:latency_vehicle}, as the number of parked vehicles increases, the processing capability of the VFs becomes more powerful. Consequently, the average system latency decreases with the growing number of vehicles in the park. DTD3 achieves lower average system latency than others as the number of vehicles increases when computing normal traffic. However, for hotspot traffic, TD3, DDPG, and PSO outperform DTD3 and SA in minimizing the average system latency. Initially, DTD3 exhibits superior performance with a smaller number of vehicles, but as the number of vehicles increases, SA surpasses DTD3 in performance.

Figure \ref{fig:utilization_vehicle} shows that the average system utilization of the system resource improves as the number of vehicles increases. DTD3 utilizes the resource more effectively than other algorithms when computing normal traffic. However, in the case of hotspot traffic, PSO and SA utilize the resource better than DTD3 until the number of vehicles reaches 200 and 400, respectively. Once the number of vehicles in VFs reaches 400, DTD3, SA, and PSO exhibit almost the same system resource utilization rate, with DTD3 slightly outperforming the others. These plots indicate that DTD3 outperforms others in both normal and hotspot traffic scenarios when there is a large number of vehicles in the VF. However, SA and PSO perform better than DTD3 when the number of vehicles in the VF is small. In contrast, TD3, DDPG, and PSO outperform SA and DTD3 in terms of average system latency in hotspot traffic scenarios.

\section{Conclusions}
In this paper, we considered the computational offloading of arrival hotspot traffic in MEC and VF networks. We formulated a multi-objective optimization problem with the objective of minimizing the average system latency and energy consumption. To find the solution to this optimization problem, we developed an equivalent RL environment representation of the MEC and VF network topology. Furthermore, we proposed a Distributed-TD3 algorithm to determine nearly optimal computational offloading decisions. The simulation results demonstrate that our approach achieves faster convergence with better performance compared to other benchmark solutions.

In the future, we plan to expand our studies to cover the optimization of offloading hotspot traffic and also investigate the influence of vehicle mobility on the completion of offloaded traffic. Additionally, we will conduct a real-world experiment to validate our findings and further enhance the robustness of our approach.




\ifCLASSOPTIONcaptionsoff
  \newpage
\fi



\bibliographystyle{IEEEtran}
%

\bibliography{DTD3}

\begin{thebibliography}{10}
\providecommand{\url}[1]{#1}
\csname url@samestyle\endcsname
\providecommand{\newblock}{\relax}
\providecommand{\bibinfo}[2]{#2}
\providecommand{\BIBentrySTDinterwordspacing}{\spaceskip=0pt\relax}
\providecommand{\BIBentryALTinterwordstretchfactor}{4}
\providecommand{\BIBentryALTinterwordspacing}{\spaceskip=\fontdimen2\font plus
\BIBentryALTinterwordstretchfactor\fontdimen3\font minus \fontdimen4\font\relax}
\providecommand{\BIBforeignlanguage}[2]{{%
\expandafter\ifx\csname l@#1\endcsname\relax
\typeout{** WARNING: IEEEtran.bst: No hyphenation pattern has been}%
\typeout{** loaded for the language `#1'. Using the pattern for}%
\typeout{** the default language instead.}%
\else
\language=\csname l@#1\endcsname
\fi
#2}}
\providecommand{\BIBdecl}{\relax}
\BIBdecl

\bibitem{intro_01}
S.~Ghazanfar, F.~Hussain, A.~U. Rehman, U.~U. Fayyaz, F.~Shahzad, and G.~A. Shah, ``Iot-flock: An open-source framework for iot traffic generation,'' in \emph{2020 International Conference on Emerging Trends in Smart Technologies (ICETST)}.\hskip 1em plus 0.5em minus 0.4em\relax IEEE, 2020, pp. 1--6.

\bibitem{intro_02}
H.~Tran-Dang and D.-S. Kim, ``Impact of task splitting on the delay performance of task offloading in the iot-enabled fog systems,'' in \emph{2021 International Conference on Information and Communication Technology Convergence (ICTC)}.\hskip 1em plus 0.5em minus 0.4em\relax IEEE, 2021, pp. 661--663.

\bibitem{intor_03}
L.~Wang, J.~Zhou, Y.~Wang, and B.~Lei, ``Energy conserved computation offloading for o-ran based iot systems,'' in \emph{ICC 2022-IEEE International Conference on Communications}.\hskip 1em plus 0.5em minus 0.4em\relax IEEE, 2022, pp. 4043--4048.

\bibitem{kar2023offloading}
B.~Kar, W.~Yahya, Y.-D. Lin, and A.~Ali, ``Offloading using traditional optimization and machine learning in federated cloud-edge-fog systems: A survey,'' \emph{IEEE Communications Surveys \& Tutorials}, vol.~25, no.~2, 2023.

\bibitem{intro_05}
P.~Porambage, J.~Okwuibe, M.~Liyanage, M.~Ylianttila, and T.~Taleb, ``Survey on multi-access edge computing for internet of things realization,'' \emph{IEEE Communications Surveys \& Tutorials}, vol.~20, no.~4, pp. 2961--2991, 2018.

\bibitem{kar2023cost}
B.~Kar, Y.-D. Lin, and Y.-C. Lai, ``Cost optimization of omnidirectional offloading in two-tier cloud--edge federated systems,'' \emph{Journal of Network and Computer Applications}, vol. 215, p. 103630, 2023.

\bibitem{intro_06}
S.~Kekki, W.~Featherstone, Y.~Fang, P.~Kuure, A.~Li, A.~Ranjan, D.~Purkayastha, F.~Jiangping, D.~Frydman, G.~Verin \emph{et~al.}, ``Mec in 5g networks,'' \emph{ETSI white paper}, vol.~28, no. 2018, pp. 1--28, 2018.

\bibitem{wu2014}
D.~Wu, Y.~Zhang, J.~Luo, and R.~Li, ``Efficient data dissemination by crowdsensing in vehicular networks,'' in \emph{2014 IEEE 22nd International Symposium of Quality of Service (IWQoS)}.\hskip 1em plus 0.5em minus 0.4em\relax IEEE, 2014, pp. 314--319.

\bibitem{zhao2017}
Y.~Zhao, Y.~Li, D.~Wu, and N.~Ge, ``Overlapping coalition formation game for resource allocation in network coding aided d2d communications,'' \emph{IEEE Transactions on Mobile Computing}, vol.~16, no.~12, pp. 3459--3472, 2017.

\bibitem{intro_07}
H.~Li, H.~Xu, C.~Zhou, X.~L{\"u}, and Z.~Han, ``Joint optimization strategy of computation offloading and resource allocation in multi-access edge computing environment,'' \emph{IEEE Transactions on Vehicular Technology}, vol.~69, no.~9, pp. 10\,214--10\,226, 2020.

\bibitem{intro_11}
W.~Yahya, E.~Oki, Y.-D. Lin, and Y.-C. Lai, ``Scaling and offloading optimization in pre-cord and post-cord multi-access edge computing,'' \emph{IEEE Transactions on Network and Service Management}, vol.~18, no.~4, pp. 4503--4516, 2021.

\bibitem{lin2023cost}
B.-S. Lin, B.~Kar, T.-L. Chin, Y.-D. Lin, and C.-Y. Chen, ``Cost optimization of cloud-edge-fog federated systems with bidirectional offloading: one-hop versus two-hop,'' \emph{Telecommunication Systems}, vol.~84, no.~4, pp. 487--505, 2023.

\bibitem{lin2019cost}
Y.-D. Lin, J.-C. Hu, B.~Kar, and L.-H. Yen, ``Cost minimization with offloading to vehicles in two-tier federated edge and vehicular-fog systems,'' in \emph{2019 IEEE 90th Vehicular Technology Conference (VTC2019-Fall)}.\hskip 1em plus 0.5em minus 0.4em\relax IEEE, 2019, pp. 1--6.

\bibitem{agbaje2022survey}
P.~Agbaje, A.~Anjum, A.~Mitra, E.~Oseghale, G.~Bloom, and H.~Olufowobi, ``Survey of interoperability challenges in the internet of vehicles,'' \emph{IEEE Transactions on Intelligent Transportation Systems}, vol.~23, no.~12, pp. 22\,838--22\,861, 2022.

\bibitem{intro_13}
B.~Kar, K.-M. Shieh, Y.-C. Lai, Y.-D. Lin, and H.-W. Ferng, ``Qos violation probability minimization in federating vehicular-fogs with cloud and edge systems,'' \emph{IEEE Transactions on Vehicular Technology}, vol.~70, no.~12, pp. 13\,270--13\,280, 2021.

\bibitem{dai2020}
P.~Dai, K.~Hu, X.~Wu, H.~Xing, F.~Teng, and Z.~Yu, ``A probabilistic approach for cooperative computation offloading in mec-assisted vehicular networks,'' \emph{IEEE Transactions on Intelligent Transportation Systems}, vol.~23, no.~2, pp. 899--911, 2020.

\bibitem{mukherjee2019}
M.~Mukherjee, S.~Kumar, C.~X. Mavromoustakis, G.~Mastorakis, R.~Matam, V.~Kumar, and Q.~Zhang, ``Latency-driven parallel task data offloading in fog computing networks for industrial applications,'' \emph{IEEE Transactions on Industrial Informatics}, vol.~16, no.~9, pp. 6050--6058, 2019.

\bibitem{intro_19}
L.-H. Yen, J.-C. Hu, Y.-D. Lin, and B.~Kar, ``Decentralized configuration protocols for low-cost offloading from multiple edges to multiple vehicular fogs,'' \emph{IEEE Transactions on Vehicular Technology}, vol.~70, no.~1, pp. 872--885, 2020.

\bibitem{rw_01}
X.~He, H.~Lu, M.~Du, Y.~Mao, and K.~Wang, ``Qoe-based task offloading with deep reinforcement learning in edge-enabled internet of vehicles,'' \emph{IEEE Transactions on Intelligent Transportation Systems}, vol.~22, no.~4, pp. 2252--2261, 2020.

\bibitem{rw_02}
W.~Zhan, C.~Luo, J.~Wang, C.~Wang, G.~Min, H.~Duan, and Q.~Zhu, ``Deep-reinforcement-learning-based offloading scheduling for vehicular edge computing,'' \emph{IEEE Internet of Things Journal}, vol.~7, no.~6, pp. 5449--5465, 2020.

\bibitem{rw_03}
Y.~Cui, L.~Du, H.~Wang, D.~Wu, and R.~Wang, ``Reinforcement learning for joint optimization of communication and computation in vehicular networks,'' \emph{IEEE Transactions on Vehicular Technology}, vol.~70, no.~12, pp. 13\,062--13\,072, 2021.

\bibitem{rw_04}
H.~Zhou, K.~Jiang, X.~Liu, X.~Li, and V.~C. Leung, ``Deep reinforcement learning for energy-efficient computation offloading in mobile-edge computing,'' \emph{IEEE Internet of Things Journal}, vol.~9, no.~2, pp. 1517--1530, 2021.

\bibitem{rw_05}
N.~Waqar, S.~A. Hassan, A.~Mahmood, K.~Dev, D.-T. Do, and M.~Gidlund, ``Computation offloading and resource allocation in mec-enabled integrated aerial-terrestrial vehicular networks: A reinforcement learning approach,'' \emph{IEEE Transactions on Intelligent Transportation Systems}, vol.~23, no.~11, pp. 21\,478--21\,491, 2022.

\bibitem{rw_06}
Y.~Liu, H.~Yu, S.~Xie, and Y.~Zhang, ``Deep reinforcement learning for offloading and resource allocation in vehicle edge computing and networks,'' \emph{IEEE Transactions on Vehicular Technology}, vol.~68, no.~11, pp. 11\,158--11\,168, 2019.

\bibitem{rw_08}
Z.~Ning, P.~Dong, X.~Wang, L.~Guo, J.~J. Rodrigues, X.~Kong, J.~Huang, and R.~Y. Kwok, ``Deep reinforcement learning for intelligent internet of vehicles: An energy-efficient computational offloading scheme,'' \emph{IEEE Transactions on Cognitive Communications and Networking}, vol.~5, no.~4, pp. 1060--1072, 2019.

\bibitem{rw_09}
Z.~Wu and D.~Yan, ``Deep reinforcement learning-based computation offloading for 5g vehicle-aware multi-access edge computing network,'' \emph{China Communications}, vol.~18, no.~11, pp. 26--41, 2021.

\bibitem{rw_10}
Y.~Xia, L.~Wu, Z.~Wang, X.~Zheng, and J.~Jin, ``Cluster-enabled cooperative scheduling based on reinforcement learning for high-mobility vehicular networks,'' \emph{IEEE Transactions on Vehicular Technology}, vol.~69, no.~11, pp. 12\,664--12\,678, 2020.

\bibitem{zhou2022}
H.~Zhou, Z.~Zhang, Y.~Wu, M.~Dong, and V.~C. Leung, ``Energy efficient joint computation offloading and service caching for mobile edge computing: A deep reinforcement learning approach,'' \emph{IEEE Transactions on Green Communications and Networking}, 2022.

\bibitem{budhiraja2022}
I.~Budhiraja, N.~Kumar, H.~Sharma, M.~Elhoseny, Y.~Lakys, and J.~J. Rodrigues, ``Latency-energy tradeoff in connected autonomous vehicles: a deep reinforcement learning scheme,'' \emph{IEEE Transactions on Intelligent Transportation Systems}, 2022.

\bibitem{sm_01}
L.~Kleinrock, \emph{Queueing systems: theory}.\hskip 1em plus 0.5em minus 0.4em\relax John Wiley, 1975.

\bibitem{ps_01}
R.~Sutton, ``Barto: Reinforcement learning: An introduction,'' \emph{IEEE Trans. Neural Netw}, vol.~9, p. 1054, 1998.

\bibitem{ps_02}
S.~Fujimoto, H.~Hoof, and D.~Meger, ``Addressing function approximation error in actor-critic methods,'' in \emph{International conference on machine learning}.\hskip 1em plus 0.5em minus 0.4em\relax PMLR, 2018, pp. 1587--1596.

\bibitem{ddpg}
T.~P. Lillicrap, J.~J. Hunt, A.~Pritzel, N.~Heess, T.~Erez, Y.~Tassa, D.~Silver, and D.~Wierstra, ``Continuous control with deep reinforcement learning,'' \emph{arXiv preprint arXiv:1509.02971}, 2015.

\bibitem{ra_01}
E.~Sergeant, ``Epitools epidemiological calculators,'' 2018.

\bibitem{Greg2016}
G.~Brockman, V.~Cheung, L.~Pettersson, J.~Schneider, J.~Schulman, J.~Tang, and W.~Zaremba, ``Openai gym,'' 2016.

\bibitem{ra_02}
A.~Raffin, A.~Hill, A.~Gleave, A.~Kanervisto, M.~Ernestus, and N.~Dormann, ``Stable-baselines3: Reliable reinforcement learning implementations,'' \emph{The Journal of Machine Learning Research}, vol.~22, no.~1, pp. 12\,348--12\,355, 2021.

\bibitem{ra_04}
S.~Kirkpatrick, C.~D. Gelatt~Jr, and M.~P. Vecchi, ``Optimization by simulated annealing,'' \emph{science}, vol. 220, no. 4598, pp. 671--680, 1983.

\bibitem{salehizadeh2009local}
S.~M.~A. Salehizadeh, P.~Yadmellat, and M.~B. Menhaj, ``Local optima avoidable particle swarm optimization,'' in \emph{2009 IEEE Swarm Intelligence Symposium}.\hskip 1em plus 0.5em minus 0.4em\relax IEEE, 2009, pp. 16--21.

\bibitem{ra_03}
T.~Akiba, S.~Sano, T.~Yanase, T.~Ohta, and M.~Koyama, ``Optuna: A next-generation hyperparameter optimization framework,'' in \emph{Proceedings of the 25th {ACM} {SIGKDD} International Conference on Knowledge Discovery and Data Mining}, 2019.

\end{thebibliography}

\end{document}